# Negative orbital Hall effect in Germanium

E. Santos,[1] J. E. Abrão,[1] J. L. Costa,[1] J. G. S. Santos,[1] J. B. S. Mendes,[2] and A. Azevedo[1]

[1]*Departamento de Física, Universidade Federal de Pernambuco, 50670-901, Recife, Pernambuco, Brazil.*

[2]*Departamento de Física, Universidade Federal de Viçosa, 36570-900, Viçosa, Minas Gerais, Brazil.*

Our investigation reveals a groundbreaking discovery of a negative inverse orbital Hall effect (IOHE) in Ge thin films. We employed the innovative orbital pumping technique where spin-orbital coupled current is injected into Ge films using YIG/Pt(2)/Ge($t_{Ge}$) and YIG/W(2)/Ge($t_{Ge}$) heterostructures. Through comprehensive analysis, we observe significant reductions in the signals generated by coherent (RF-driven) and incoherent (thermal-driven) spin-orbital pumping techniques. These reductions are attributed to the presence of a remarkable strong negative IOHE in Ge, showing its magnitude comparable to the spin-to-charge signal in Pt. Our findings reveal that although the spin-to-charge conversion in Ge is negligible, the orbital-to-charge conversion exhibits large magnitude. Our results are innovative and pioneering in the investigation of negative IOHE by the injection of spin-orbital currents.

The orbital Hall effect (OHE) occurs when there is a transverse flow of orbital angular momentum (OAM) induced by an external electric field [1-4], like the spin Hall effect (SHE) [5-7]. Notably, the OHE operates independently of the existence of spin-orbit coupling (SOC). Recent studies [4, 8] have revealed substantial negative values for orbital Hall conductivity ($\sigma_{OH}$) across different materials. However, these materials often exhibit significant spin Hall conductivity ($\sigma_{SH}$) presenting challenges in isolating distinct spin and orbital contributions. Addressing this complexity, our understanding of OHE is evolving, with persistent investigations using various strategies aimed at unraveling the intrinsic and extrinsic mechanisms underlying this phenomenon. Exploration of the intrinsic and extrinsic mechanisms governing the OHE has encompassed different classes of materials, from transition metals and semiconductors to two-dimensional materials and topological insulators [9-19]. Despite this extensive investigation, there has been a notable absence of experimental studies investigating semiconductors materials. This gap in knowledge is particularly intriguing given the potential of group IV semiconductors, such as Ge and Si, to serve as exceptional platforms for spintronics phenomena [20-23]. Ge has a much higher carrier mobility than Si, which can be used to improve the performance of transistors based on this material. Currently, Ge has applications in optical fibers and optical tweezers, while Si-Ge alloys play a role in microchip manufacturing, with feature sizes on the chips reaching 7 nm [24-27]. This unique combination of properties establishes group IV semiconductors as attractive for fundamental research in OHE and practical advances in spin-orbitronics applications.

In turn, a recent theoretical work [28] has discovered that Ge has $\sigma_{OH}^{Ge} \sim -1270 (\hbar/e)(\Omega \cdot \text{cm})^{-1}$ and $\sigma_{SH}^{Ge} \sim 1.6 \times 10^{-1} (\hbar/e)(\Omega \cdot \text{cm})^{-1}$, making it a prime material for studying orbital effects. Materials with negative $\sigma_{OH}$, like Ge, play a crucial role in distinguishing orbital from spin effects and offer insights for the development of new OAM-based devices. Nonetheless, to our knowledge, the OHE in Ge has not been investigated to date. In this work, we experimentally investigate orbital-charge conversion in Ge thin films using the inverse orbital Hall effect. The measurements were carried out at room temperature using the spin pumping driven by ferromagnetic resonance (SP-FMR) and longitudinal spin Seebeck effect (LSSE) techniques. We fabricated samples of YIG/Pt, YIG/Ge, YIG/Pt/Ge, and YIG/W/Ge, where YIG refers to yttrium iron garnet ($Y_3Fe_5O_{12}$) grown on (111)-oriented Gadolinium Gallium Garnet ($Gd_3Ga_5O_{12}$, GGG), by liquid phase epitaxy (LPE) through the traditional $PbO/B_2O_3$ flux method. The quality of the YIG samples is attested by the small FMR linewidth, which is less than 1 Oe (see figure 1(b)). All other thin films were deposited using magnetron sputtering at room temperature, with a working pressure of 2.8 mTorr and a base pressure of $1.5 \times 10^{-7}$ Torr. All investigated samples have lateral dimensions of 3.0 x 1.5 mm.

Figure 1(a) illustrates the SP-FMR process in a YIG/Pt bilayer. The precessing magnetization injects a spin current density into the Pt layer, given by $\vec{J}_S = (\hbar g_{eff}^{\uparrow\downarrow}/4\pi M^2)(\vec{M} \times \dot{\vec{M}})$ [29]. This spin current manifests itself as two components: an AC represented by the orange vectors and a DC represented by the fixed red arrow parallel to the applied field. The DC component induces a spin accumulation that diffuses into the Pt layer, characterized by spin polarization ($\hat{\sigma}_S$) oriented along the z-axis with characteristic diffusion length, typically spanning a few nanometers in materials with large SOC [10]. Figures 1 (b-d) shows the derivatives of the FMR absorption curves with the external field applied in-plane for bare YIG(400), YIG(400)/Ge(8) and YIG(400)/Pt(8). The values in parentheses denote the film thicknesses in nanometers. To excite the FMR condition, the samples were placed at the bottom of a rectangular microwave resonant cavity operating at 9.41 GHz, with an incident power of 15 mW. For more details on the SP-FMR technique, refer to [10, 30]. The experimental data were fitted using the derivative of a Lorentzian curve. The numerical fit yields FMR linewidths of $\Delta H = 0.89\ Oe$, $\Delta H = 0.90\ Oe$ and $\Delta H = 2.00\ Oe$, for bare YIG, YIG/Ge(8) and YIG/Pt(8), respectively. The spin pumping effect in ferromagnetic(FM)/normal-metal(NM) introduces additional damping in the FMR process, meaning that an extra term is added to the Landau-Lifshitz-Gilbert equation [29]. The increase in FMR linewidth is described by $\Delta H_{SP} = \hbar \omega_0 \, g_{eff}^{\uparrow\downarrow}/4\pi t_{FM} M$, where $g_{eff}^{\uparrow\downarrow}$ is the real part of the effective spin mixing conductance, $\omega_0$ is the angular frequency, $t_{FM}$ and $M$ are the thickness and magnetization of the FM, respectively. Comparing figures 1(b) and 1(c), we observe practically no change in linewidth. As expected, due to the large SOC of Pt, the FMR linewidth of YIG/Pt(8) in figure 1(d) exhibited a large increase to 2 Oe, compared with bare YIG. Figure 1(e) shows the numerical fits of the three bilayers, where the increase of $\Delta H$ in the YIG/Pt(8) bilayer due to the SP process is highlighted.

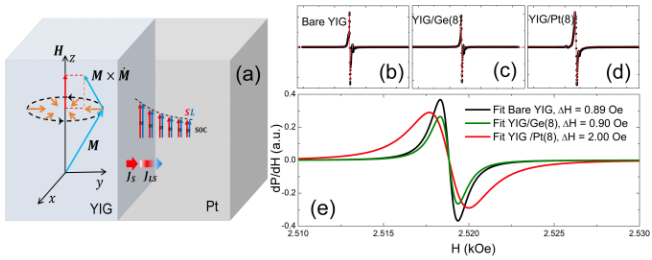
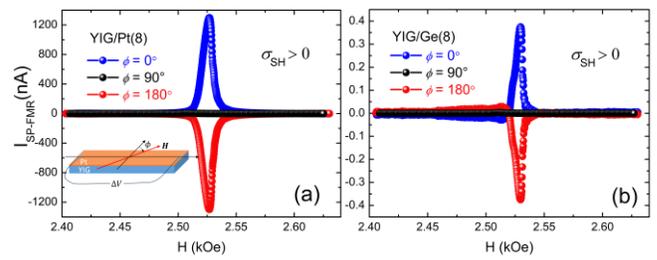

FIG. 1. (a) Illustration of the SP-FMR process where the red arrows represent the DC component of the spin current $\vec{J}_S \propto (\vec{M} \times \dot{\vec{M}})$ that diffuses into the Pt layer, where $\vec{J}_S$ transform into $\vec{J}_{LS}$ due to SOC. (b-d) FMR absorption curves (black symbols) for: bare YIG(400), YIG(400)/Ge(8), and YIG(400)/Pt(8), respectively. The solid red lines were obtained by fitting the data with derivative of a Lorentzian curve. The weak absorption peaks that appear below and above the resonance field of the uniform mode are due to surface and volume magnetostatic modes. (e) shows the solid curves obtained by the numerical fits, where the large increase in $\Delta H$ of YIG/Pt(8) is highlighted.

Subsequently, SP-FMR measurements were conducted by attaching electrodes at the edges of the metal layers using silver paint. All subsequent SP-FMR measurements were obtained at the same frequency of 9.41 GHz and using an incident power of 43 mW. Figures 2 (a) and 2(b) illustrate the electrical current generated by the SP-FMR process as function of the external applied field, for YIG/Pt(8) and YIG/Ge(8), respectively. The electrical current is defined as $I_{NM} = V/R_S$, where $V$ is the electrical voltage that is directly measured via a nanovoltmeter and $R_S$ the electrical sheet resistance of the film. As illustrated in figure 1(a), the DC spin current, represented by the red arrows and possessing spin polarization $\hat{\sigma}_S$ along z-axis, diffuses into the Pt layer. The large SOC in Pt leads to the coupling of spin and orbital states [9-11,31]. Consequently, the spin-orbital current along the $\hat{y}$ direction is expressed as $\vec{J}_S(y) = A \frac{\sinh[(t_{NM}-y)/\lambda_1]}{\sinh(t_{NM}/\lambda_1)} \hat{y} + B \frac{\sinh[(t_{NM}-y)/\lambda_2]}{\sinh(t_{NM}/\lambda_2)} \hat{y}$. Here, the constants A and B are to be determined via boundary conditions, $\lambda_1$ and $\lambda_2$ are diffusion lengths, that depend on both the spin diffusion length $\lambda_S$, orbital diffusion length $\lambda_L$, and diffusion coupling parameter $\lambda_{LS}$ [11, 31]. Conversion of spin current to charge current occurs due to the inverse spin Hall effect (ISHE) [32,33]. The relationship between spin current ($\vec{J}_S$) and charge current ($\vec{J}_C$), within the NM, is given by $\vec{J}_C = \theta_{SH}(\hat{\sigma}_S \times \vec{J}_S)$, where $\theta_{SH} = (2e/\hbar)\sigma_{SH}/\sigma_e$ is the spin Hall angle, representing the spin-charge conversion efficiency and $\sigma_e$ is the electric conductivity. The orbital current generated by the coupling $\vec{L} \cdot \vec{S}$ is $\vec{J}_L(y) = C\delta_{LS}\vec{J}_S(y)$, where the dimensionless constant C represents the strength of this relationship, and $\delta_{LS} = \pm 1$ indicates the SOC signal. The conversion of orbital current to charge current occurs due to the IOHE [1,10,34,35]. Analogously to ISHE, we can write the mathematical relationship between the orbital current and the charge current as $\vec{J}_C = \theta_{OH}(\hat{\sigma}_L \times \vec{J}_L)$, where $\theta_{OH} = (2e/\hbar)\sigma_{OH}/\sigma_e$ is the orbital Hall angle and $\hat{\sigma}_L$ is the orbital polarization, that couples parallel (positive SOC) or antiparallel (negative SOC) to $\hat{\sigma}_S$. The resultant charge current is the cumulative effect of currents generated through both the ISHE and IOHE, denoted as $\vec{J}_C^{eff} = \vec{J}_C^{ISHE} + \vec{J}_C^{IOHE}$. It is noteworthy that despite these effects describing similar phenomena, their physical origins are distinct [1,6,8,10]. Furthermore, the polarity of the SP-FMR signal is determined by the strength of spin orbit coupling in the material. Figure 2 shows the results of SP-FMR measurements conducted on two distinct samples: YIG/Pt(8) and YIG/Ge(8) at $\phi = 0°$ (blue symbols), $\phi = 180°$ (red symbols), and $\phi = 90°$ (black symbols), where $\phi$ is defined in the inset of figure 2(a). As Pt has strong SOC, we expect large spin and orbital currents represented by $\vec{J}_S$ and $\vec{J}_L$. The spin Hall orbital Hall conductivities for Pt are denoted as $\sigma_{SH}^{Pt} \sim 2012 (\hbar/e)(\Omega \cdot cm)^{-1}$ and $\sigma_{OH}^{Pt} \sim 144 (\hbar/e)(\Omega \cdot cm)^{-1}$[8], respectively. Consequently, the dominant contribution to the charge current produced via SP-FMR in YIG/Pt(8) is primarily due to ISHE. Note that the SP-FMR signals in figure 2 obey the ISHE equation. A peak with positive polarity is observed for $\phi = 0°$, while the signal changes its polarity for $\phi = 180°$ and at $\phi = 90°$ the measured signal is null. The substantial SP-FMR signal observed in Pt (figure 2(a)) contrasts with the weak signal in Ge (figure 2(b)), attributed to its weak SOC. The ratio between the SP-FMR signals generated in YIG/Ge(8) and the one generated in YIG/Pt(8) is calculated as, $I_{YIG/Ge(8)}^{Peak}/I_{YIG/Pt(8)}^{Peak} \sim 2.5 \times 10^{-4}$. Furthermore, the negligible SOC in Ge results in scarce orbital current generation during spin current propagation within the material.

FIG. 2. SP-FMR signals for (a) YIG/Pt(8), and (b) YIG/Ge(8). Both signals obey the ISHE equation, $\vec{J}_C = \theta_{SH}(\hat{\sigma}_S \times \vec{J}_S)$. Due to the weak strength of SOC in Ge, the ISHE signal is significantly decreased compared to the Pt signal, while both materials exhibit $\sigma_{SH} > 0$. Inset of figure (a) defines the angle $\phi$.

In the next step of our work, we studied YIG/Pt(2)/Ge heterostructures, where the YIG/Pt(2) bilayer is used to inject an orbital current into Ge films. Notably, in the previous YIG/Ge configuration, the SP process injects only pure spin current in Ge. The strong SOC of Pt couples $\vec{L}$ and $\vec{S}$, resulting in an entangled current $\vec{J}_{LS}$. The current $\vec{J}_{LS}$ reaches Ge layer, where exclusively the orbital current is converted into charge current $J_C^{Ge}$ through the IOHE. In YIG/Pt(2)/Ge, the effective charge current is $\vec{J}_C^{eff} = (2e/\hbar)\theta_{SH}^{Pt}(\hat{\sigma}_S \times \vec{J}_S^{Pt}) + (2e/\hbar)\theta_{OH}^{Ge}(\hat{\sigma}_L \times \vec{J}_L^{Ge})$, where $\hat{\sigma}_L = \hat{\sigma}_S = \hat{z}$, $\vec{J}_S = J_S\hat{y}$, $\vec{J}_L = J_L\hat{y}$, $\theta_{SH}^{Pt} > 0$ and $\theta_{OH}^{Ge} < 0$. Analyzing the equation for $\vec{J}_C^{eff}$, it becomes clear that the first term generates a current along $+\hat{x}$ direction, while the second term is along the $-\hat{x}$ direction. This indicates a reduction in the measured voltage value. The effective charge current generated in YIG/Pt/Ge $\vec{J}_C^{Ge}$ is obtained by the subtraction between the $\vec{J}_C^{eff}$ and $\vec{J}_C^{Pt(2)}$. To better understand this behavior, we fabricated a series of YIG/Pt(2)/Ge($t_{Ge}$) samples, varying the thickness of the Ge layer from 2 nm to 50 nm, where we analyze $I_{YIG/Pt(2)/Ge}^{Peak}$. In figure 3(a), when $t_{Ge} = 0$ nm, $I_{YIG/Pt(2)}^{Peak}$ reaches a maximum of approximately 600 nA, signifying no contribution from the Ge layer. With $t_{Ge} = 2$ nm, $I_{YIG/Pt(2)/Ge(2)}^{Peak}$ is around 370 nA, revealing that only 2 nm of Ge is sufficient to induce a negative orbital-charge conversion, equivalent to roughly 60% of the ISHE in YIG/Pt(2). As the thickness of the Ge layer is progressively increased, a consistent decline in the signal becomes evident. For example, at $t_{Ge} = 10$ nm, $I_{YIG/Pt(2)/Ge(10)}^{Peak}$ is approximately 140 nA indicating a reduction of 76% of the ISHE in YIG/Pt(2). The experimental data saturates for $t_{Ge} > 30$ nm, where $I_{YIG/Pt(2)/Ge(30)}^{Peak} \sim 0$. Therefore, at the saturation, $|I_{IOHE}^{Ge}| \sim I_{ISHE}^{Pt}$. In $\phi = 0°$, as shown by the blue symbols, we have $\hat{\sigma}_S \| \hat{\sigma}_L \| \vec{H}$. Upon inverting the external magnetic field $\vec{H}$ for $\phi = 180°$, illustrated in figure 3(a) by the red symbols, $\hat{\sigma}_S$ and $\hat{\sigma}_L$ invert direction while remaining parallel, owing to the strong SOC of Pt. By inverting $\hat{\sigma}_S$ we expect a negative ISHE in Pt and a positive IOHE signal in Ge, according to the respective ISHE and IOHE equations. Consequently, at $\phi = 180°$, the SP-FMR signal tends towards zero in a similar way. The inset in figure 3(a) illustrates the physical scheme of the effective charge current in YIG/Pt/Ge, measured by SP-FMR for $\phi = 0°$ and $\phi = 180°$.

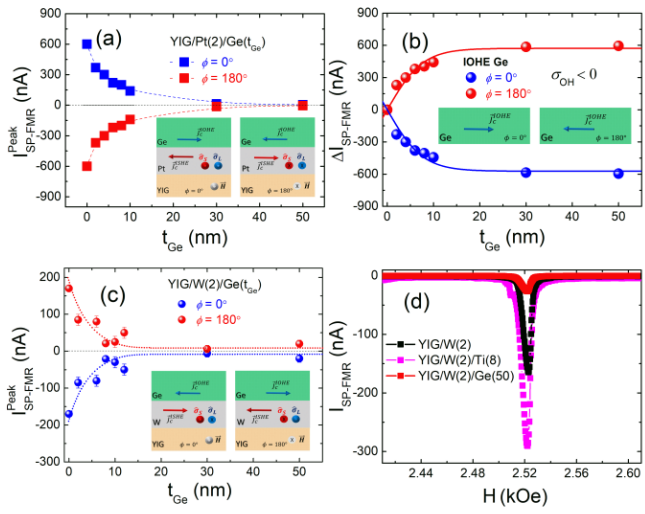
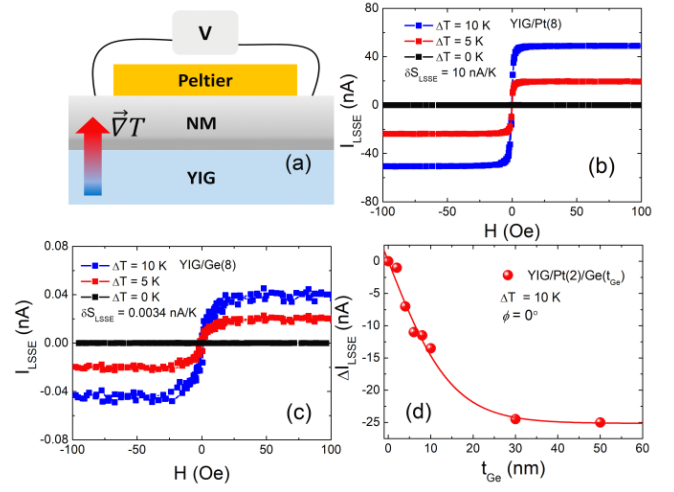

FIG. 3. (a) Ge thickness dependence of the SP-FMR peak signal of YIG/Pt(2)/Ge($t_{Ge}$). Blue data corresponds to $\phi = 0°$ and red data corresponds to $\phi = 180°$. Due to the negative IOHE of Ge, a gradual reduction in the signal is observed with increasing the thickness of the Ge layer. (b) IOHE signal for Ge as a function of thickness, with theoretical fit using $I_{IOHE}^{Ge} = D\tanh(t_{Ge}/2\lambda_L^{Ge})$, where we found $\lambda_L^{Ge} = (4.0 \pm 0.6)$ nm, and $\Delta I_{SP-FMR} = I_{Pt(2)} - I_{Pt(2)/Ge}$. We observed that the greater the thickness of Ge, the more intense the IOHE signal, which in magnitude is approximately equal to the ISHE of Pt(2) for $t_{Ge} > 30$ nm. The inset in (a) illustrates the physical scheme of the effective charge current in YIG/Pt/Ge, measured by SP-FMR for $\phi = 0°$ and $\phi = 180°$.

Figure 3(b) shows the IOHE signals for Ge films ranging in thickness from 0 to 50 nm, for $\phi = 0°$ (blue symbols) and $\phi = 180°$ (red symbols). These signals were obtained by calculating the difference between $I_{YIG/Pt(2)/Ge} - I_{YIG/Pt(2)}$. Given that Ge exhibits negligible SOC, we can distinctly analyze the spin and orbital contributions of the $\vec{J}_{LS}$ current that reaches the Ge layer. As confirmed in our previous result, the spin component is nearly null, highlighting the importance of the orbital component in this context. The orbital flow within the Ge layer has a well-defined orbital diffusion length $\lambda_L^{Ge}$. Using $I_{IOHE}^{Ge} = D\tanh(t_{Ge}/2\lambda_L^{Ge})$ [9-11] it is possible to estimate $\lambda_L^{Ge}$ in the YIG/Pt(2)/Ge heterostructures, where D is constant. Fitting the experimental data in figure 3(b), we found $\lambda_L^{Ge} = (4.0 \pm 0.6)$ nm. To contextualize our result concerning $\lambda_L^{Ge}$, we will discuss some details of the spin and orbital diffusion lengths. A fundamental distinction exists between spin and orbital transport. Contrary to intuition, the crystal field does not quench nonequilibrium OAM as effectively as it suppresses equilibrium OAM [31]. This is attributed to the presence of degenerate orbital states that play a crucial role in long-range orbital transport [36]. Orbital degeneracy is generally protected against crystal field splitting, allowing orbital momentum to traverse longer distances compared its spin counterpart. However, in materials with weak SOC, long spin diffusion lengths are expected [37-40]. For example, in Ge is expected spin diffusion lengths of order of micrometers [37-40]. In our experiment, a simultaneous injection of both spin current and orbital current occurs, represented by the coupled current $\vec{J}_{LS}$. The distinction between spin and orbital contributions relies on the knowledge of $\sigma_{SH}$ and $\sigma_{OH}$. Notably, in Ge, the IOHE signals are expected be much larger than the ISHE signals [28]. Thus, we can conclude that the diffusion length found is not predominantly associated with spin. On the contrary, the Orbital diffusion lengths in Ge could potentially exceed what our experiments reveal, as we do not directly inject an orbital current into the Ge layer. Initially, the orbital current is generated within the Pt layer. Prior to reaching the Ge layer, this orbital current must flow through the Pt film. Along this pathway, the strong SOC of Pt imposes constraints on both spin and orbital diffusion lengths, where $\lambda_S^{Pt} \sim 1.5$ nm or even shorter [10]. Consequently, a significant reduction in the orbital current occurs within the initial 2 nm before reaching the Ge layer. Additionally, the resistivity mismatch at the Pt(2)/Ge interface further reduces its magnitude. This phenomenon can explain the orbital diffusion length observed in our experiments.

FIG. 4. (a) Schematically shows the LSSE configuration. LSSE measurement for (b) YIG/Pt8, (c) YIG/Ge(8), and (d) IOHE for Ge films, where the orbital current was injected from YIG/Pt(2), where $\Delta I_{LSSE} = I_{Pt(2)} - I_{Pt(2)/Ge}$. Theoretical fit using $I_{IOHE}^{Ge} = A\tanh(t_{Ge}/2\lambda_L^{Ge})$, where we found $\lambda_L^{Ge} = (7.5 \pm 0.5)$ nm.

Another notable result is presented in figures 3(c) and 3(d). We successfully replicate results similar to those in figure 3(a), employing W instead of Pt. We fabricated the following samples: YIG/W(2), YIG/W(2)/Ge($t_{Ge}$), and YIG/W(2)/Ti(8). In figure 3(c), the results for YIG/W(2)/Ge($t_{Ge}$) are shown for $\phi = 0°$ and $\phi = 180°$. Figure 3(d) shows the signal for YIG/W(2), with $I_{SP-FMR}^{Peak} \sim -170$ nA. Upon the addition of Ge(50) to the YIG/W(2) sample, $I_{SP-FMR}^{Peak}$ was reduced to around $-25$ nA. Although a reduction in signal magnitude was observed, it was considerably less significant than in samples utilizing Pt. On the other hand, figure 3(d) also shows the signal for YIG/W(2)/Ti(8) in comparison to YIG/W(2). An increase in the signal by nearly a factor of 2 was observed.

It is known that W has $\sigma_{SH}^W = -768\,(\hbar/e)(\Omega \cdot cm)^{-1}$ and $\sigma_{OH}^W = 4664\,(\hbar/e)(\Omega \cdot cm)^{-1}$ [8]. Consequently, an SP-FMR signal is expected to originate from spin-orbital to charge conversion via both ISHE and IOHE, with the latter having a significantly greater magnitude. The effective charge current in YIG/W/Ge is given by $\vec{J}_C^{eff} = (2e/\hbar)[\theta_{SH}^W(\hat{\sigma}_S \times \vec{J}_S^W) + \theta_{OH}^W(\hat{\sigma}_L \times \vec{J}_L^W) + \theta_{OH}^{Ge}(\hat{\sigma}_L \times \vec{J}_L^{Ge})]$, where $\theta_{SH}^W < 0, \theta_{OH}^W > 0, \theta_{OH}^{Ge} < 0$, $\hat{\sigma}_L = -\hat{z}, \hat{\sigma}_S = \hat{z}, \vec{J}_S = J_S \hat{y}$ and $\vec{J}_L = J_L \hat{y}$. Upon analyzing the equation for $\vec{J}_C^{eff}$ we observed that first and second terms contribute $-\hat{x}$ direction, while the third term contributes in the $+\hat{x}$, resulting in a reduction in the signal. However, we observed a less substantial reduction in the signal in the sample with W compared to the one with Pt. This discrepancy can be attributed to intrinsic characteristics of W and Pt. In a preliminary analysis, we can state the following: (i) W has a smaller SOC than Pt and a large electrical resistivity [41, 42], resulting in a lower $\vec{J}_{LS}$ current. (ii) The large value of $\sigma_{OH}^W$ leads to a more pronounced orbital-charge conversion in W compared Pt, causing the residual orbital current reaching the Ge layer to be decreased.

The fluctuations observed in the data of figure 3(c) are linked to variations in the spin conductivity [42] arising from the coexistence of α and β phases of W. The simultaneous presence of α and β phases in W thin films is a relatively common occurrence in those produced through sputtering [43]. In the β phase ($t_W < 10$ nm) the resistivity of the films is very high, while in the α phase it decreases considerably [43]. Despite this challenge, a consistent reduction in the signal is evident in figure 3(c), with the dashed line serving as a visual guide. It is worth noting that the reduction in signal measured in W/Ge films follows a similar trend to that IOHE of the Ge layer. Figure 3(d) corroborates our theoretical hypotheses: (i) spin current injection into the YIG/NM bilayers results in the accumulation of spins and, due to the strong SOC of the NM, generates a collinear orbital current. (ii) In materials with negative SOC, the orbital polarization is antiparallel to the spin polarization. Note that the introduction of a Ti film on top of the YIG/W bilayer resulted in a gain of the signal. This increase can be exclusively attributed to orbital currents, given that Ti exhibits negligible SOC [11]. The reversal of orbital polarity within W, of the $\vec{J}_{LS}$ current that reaches the Ti film, generates an orbital-charge conversion in the same direction (negative) as the YIG/W bulk signal. This effect significantly enhances the SP-FMR signal. Interestingly, this increase was not observed in Ge, where the negative $\sigma_{OH}$, along with the inversion of orbital polarity, leads to a positive orbital-charge conversion due to IOHE in Ge films. This conversion occurs in the opposite direction to the bulk signal of W. Finally, the shorter orbital diffusion length can be attributed to the large resistivity of the beta-phase W films, which quickly dissipates the $\vec{J}_{LS}$ current. Hence, for a heterostructure works as an effective orbital-current injector, it is essential to employ a NM with strong SOC, like Pt, in the fabrication of YIG/NM structures. Nonetheless, the presence of highly resistive phases (like β-W) may impede the successful injection of orbital currents into adjacent films.

We also employed the LSSE technique to excite spin currents and induce orbital currents in YIG/Pt(2)/Ge($t_{Ge}$). LSSE consists of applying a thermal gradient to generate spin currents from the magnon flow generated in the YIG bulk [44], as illustrated in figure 4(a). To create the temperature gradient, we utilized a Peltier module, and the resulting temperature difference ($\Delta T$) between the bottom and top of the sample was measured using a differential thermocouple. The IOHE voltage due to LSSE was detected between the two silver-painted electrodes positioned at the edges of the Pt film. The underlying physical mechanism is similar to SP-FMR, with $I_{LSSE} = \delta S_{LSSE} \nabla T$, and $\delta S_{LSSE}$ is the Seebeck-like spin coefficient, including contributions from spin and/or orbital effects.

We investigated the behavior of the DC electric current ($I_{LSSE} = V_{LSSE}/R_S$) arising from the IOHE in response to the sweep of the external magnetic field. Figure 4(b) shows the LSSE signals for YIG/Pt(8) and figure 4(c) shows the LSSE signals for YIG/Ge(8) under temperature differences of $\Delta T = 0K$ (black), $\Delta T = 5K$ (red) and $\Delta T = 10K$ (blue). We observed a very small YIG/Ge(8) signal, when compared to the YIG/Pt(8) signal, similar to the trend observed in SP-FMR. Figure 4(d) shows the IOHE for Ge, and through fitting the IOHE signal, we obtained $\lambda_L^{Ge} = (7.5 \pm 0.5)$ nm. Furthermore, there is a difference in the diffusion length measured by SP-FMR and LSSE, which has already been discussed in [9-11]. Although both processes (SP-FMR and LSSE) are of spin current injection, there is a basic difference between the two processes. While in SP-FMR the spin injection is interfacial, in LSSE the spin injection is due to the magnon current generated in the YIG bulk.

In conclusion, our work investigates the IOHE in Ge films. Although Ge has negligible SOC, it manifests a substantial and negative IOHE value, comparable in magnitude to the ISHE signal observed in Pt. Efficient injection of spin-orbital currents has been achieved through of heterostructures, specifically YIG(400)/HM(2), where the heavy metal (HM) layer could be Pt or W. Through a careful combination of different stack layers and considering the spin and orbital conductivities, we elucidate successfully the exciting results obtained through the SP-FMR and LSSE techniques. Our study highlights the critical role of orbital polarization in influencing IOHE, a factor that can be controlled through the heavy metal SOC signal. By exploring materials with a prominent IOHE and a negligible ISHE, we effectively isolate and distinguish spin effects from orbital effects. These discoveries not only contribute valuable insights to the field of orbitronics, but also have potential applications in the development of electronic devices based on orbital angular momentum flow and spin-orbital coupling.

This research is supported by Conselho Nacional de Desenvolvimento Científico e Tecnológico (CNPq), Coordenação de Aperfeiçoamento de Pessoal de Nível Superior (CAPES) (Grant No. 0041/2022), Financiadora de Estudos e Projetos (FINEP), Fundação de Amparo à Ciência e Tecnologia do Estado de Pernambuco (FACEPE), Universidade Federal de Pernambuco, Multiuser Laboratory

---